\documentclass[aps,pra,twocolumn,superscriptaddress,10pt]{revtex4-1}
\usepackage[utf8]{inputenc}
\usepackage[T1]{fontenc}
\usepackage{graphicx}

\usepackage{dcolumn}
\usepackage{bm}
\usepackage{multirow}
\usepackage{amssymb}
\usepackage{float}
\usepackage{amsfonts}
\usepackage{amsmath}
\usepackage{amsthm}
\usepackage{enumerate}
\usepackage{tikz}
\usetikzlibrary{shapes,arrows}
\usepackage{mathtools}
\usepackage{xcolor}
\usepackage{color}
\usepackage{comment}
\usepackage{marvosym}
\usepackage{braket}

\usepackage[pdftex,
pdfauthor={J. Carrasco et al.},
pdftitle={Theoretical and Experimental Perspectives of Quantum Verification}
colorlinks=true,
allcolors=black,hidelinks]{hyperref}

\newcommand{\ketbra}[2]{\ket{#1}\bra{#2}}

\newcommand{\bi}{\begin{itemize}}
\newcommand{\ei}{\end{itemize}}

\newcommand{\be}{\begin{equation}}
\newcommand{\ee}{\end{equation}}
\newcommand{\bea}{\begin{eqnarray}}
\newcommand{\eea}{\end{eqnarray}}

\usepackage[most]{tcolorbox}
\usepackage{ragged2e}
\newtcolorbox[auto counter]{mybox}[2][]{
	attach boxed title to top center= {yshift=-8pt},
	colback      = blue!35!black!5!white,
	colframe     = blue!35!black,
	fonttitle    = \bfseries\sffamily,
	colbacktitle = blue!35!black,
	title        = {Box \thetcbcounter: #2},
	enhanced,
	#1
}

\newcommand*{\trAE}[2][]{\ensuremath{\textrm{Tr}_{#1}\left[ #2 \right]}}

\begin{document}

\title{Theoretical and Experimental Perspectives of Quantum
  Verification}%

\date{\today}

\author{Jose Carrasco} \affiliation{Institute for Theoretical Physics,
  University of Innsbruck, Austria} \author{Andreas Elben}
\affiliation{Center for Quantum Physics, University of Innsbruck,
  Austria} \affiliation{Institute for Quantum Optics and Quantum
  Information of the Austrian Academy of Sciences, Innsbruck, Austria}
\author{Christian Kokail} \affiliation{Center for Quantum Physics,
  University of Innsbruck, Austria} \affiliation{Institute for Quantum
  Optics and Quantum Information of the Austrian Academy of Sciences,
  Innsbruck, Austria} \author{Barbara Kraus} \affiliation{Institute
  for Theoretical Physics, University of Innsbruck, Austria}
\author{Peter Zoller} \affiliation{Center for Quantum Physics,
  University of Innsbruck, Austria} \affiliation{Institute for Quantum
  Optics and Quantum Information of the Austrian Academy of Sciences,
  Innsbruck, Austria}

\begin{abstract}
  In this perspective we discuss verification of quantum devices in
  the context of specific examples, formulated as proposed
  experiments. Our first example is verification of analog quantum
  simulators as Hamiltonian learning, where the input Hamiltonian as
  design goal is compared with the parent Hamiltonian for the quantum
  states prepared on the device. The second example discusses
  cross-device verification on the quantum level, i.e.~by comparing
  quantum states prepared on different quantum devices. We focus in
  particular on protocols using randomized measurements, and we
  propose establishing a central data repository, where existing
  experimental devices and platforms can be compared. In our final
  example, we address verification of the output of a quantum device
  from a computer science perspective, addressing the question of how
  a user of a quantum processor can be certain about the correctness
  of its output, and propose minimal demonstrations on present day
  devices.
\end{abstract}

\maketitle

\section{Introduction}\label{sec:level1}

The dream and vision of now more than two decades to build quantum
computers and quantum simulators has materialized as nascent
programmable quantum devices in today's
laboratories~\cite{Deutsch2020-rw,preskill2018quantum,NAP25613}. While
first generation experiments focused on basic demonstration of
building blocks of quantum information processing, quantum
laboratories now host programmable intermediate scale quantum devices,
which - while still imperfect and noisy - open the perspective of
building quantum machines, which fulfill the promise of becoming more
powerful than their classical counterparts. Significant advances in
building small scale quantum computers and quantum simulators have
been reported with various physical platforms, from atomic and
photonic systems to solid state devices. A central aspect in further
developments is verification of proper functioning of these quantum
devices, including cross-device and cross-platform
verification. Quantum verification is particularly challenging in
regimes where comparison with classical simulation of quantum devices
is no longer feasible.

Quantum characterization, validation and verification (QCVV) is a
well-developed field in quantum information theory, and we refer to
reviews~\cite{eisert2020quantum,GhKa19,Supic2020} and
tutorials~\cite{Kliesch2020-cy} on this topic. The challenge in
designing practical techniques to characterize quantum processes on
intermediate and large-scale quantum devices is related to the (in
general) exponential scaling of number of experiments and digital
post-processing resources with system size, as is manifest in quantum
process tomography or state tomography.  Exponential resources can be
circumvented by extracting partial information about quantum processes
providing a figure of merit, such as a process fidelity. However, such
protocols also face the requirement of decoupling the state
preparation and measurement errors from a process
fidelity. Applications of well established protocols in experimental
settings, for example as randomized or cycle benchmarking of quantum
computers~\cite{erhard2019characterizing} or verifiable
measurement-based quantum computation~\cite{BFKW13} have been
reported.

In this `perspective' we wish to look forward to possible near future
experiments addressing verification of quantum computers and quantum
simulators, and in particular venturing into less explored
territories. We illustrate aspects of verification, which are
physically relevant and conceptually complementary to previous work,
by describing three experimental scenarios as `proposed
experiments'. Our discussion aims at connecting recent theoretical
results with possible implementation of verification protocols in
existing experimental settings. Clearly, different communities from
quantum experimentalists to theorists, and computer scientists look at
perspectives on verification from quite different angles, and our
examples are chosen to reflect this diversity.

Our first example illustrates verification of analog quantum
simulators~\cite{cirac2012goals,NAP25613} via \textit{Hamiltonian
  learning}~\cite{bairey2019learning,qi2019determining,li2020hamiltonian}. The
central idea is to verify the analog quantum simulator by comparing
the desired many-body Hamiltonian, i.e.~the Hamiltonian to be
implemented, with the actual, physically realized Hamiltonian, which
can be efficiently reconstructed from measurements of quantum states
prepared on the quantum device. This is applicable to, and immediately
relevant for present analog quantum simulation experiments for spin
and Hubbard models with atoms and ions, and superconducting
qubits~\cite{altman2019quantum,tarruell2018quantum,browaeys2020many,morgado2020quantum,ebadi2020quantum,mazurenko2017cold,arguello2019analogue,schafer2020tools,Kokail2019,monroe2019programmable,huang2020superconducting}.

In our second example we address cross-device and cross-platform
verification as applicable to quantum computers and quantum
simulators. Here the goal is the pairwise \textit{comparison of
  quantum states} implemented on different quantum devices on the
level of the full many-qubit wave function, or for reduced density
matrices of subsystems. To this end, results of randomized
measurements, performed on each device separately, can be classically
correlated to estimate the fidelity of two quantum states, with
efficiency scaling better with (sub-) system size than what is
achieved in quantum state tomography~\cite{Elben2020,Flammia2020}. We
envision a community effort where data from randomized measurements
are uploaded to a central data repository, enabling the direct
comparison of multiple quantum devices for a defined set of quantum
problems, specified either as quantum circuits and algorithms or
Hamiltonian evolution.

Finally, in our third example we move on to verification from a
computer scientist perspective, and address the question of how a user
of a quantum processor can be certain about the correctness of its
output. This question becomes particularly important in case the user
of a quantum device does not have direct access to it (e.g. cloud
computing). Is it even possible for a user to rely on the result if
they cannot verify it efficiently themselves? This question has been
answered in the affirmative in case the user has access to a limited
amount of quantum
resources~\cite{MF13,Mo14,HM15,GKP15,HPF15,FK17,ABEM17,TMH19}. Interestingly,
such a \textit{verification of the output} is feasible even via purely
classical means~\cite{Ma18}. However, not very surprisingly, the
resources required to implement such a verification protocol are
beyond reach with current technology. Due to the rapid technological
developments and the accompanying need for the ability to verify the
output of a computation, we propose here a proof-of-principle
experiment to implement such a verification protocol that is feasible
with current technologies.

\section{Verification of Analog Quantum Simulators via Hamiltonian
  Learning}

The goal of quantum simulation is to solve the quantum many-body
problem~\cite{cirac2012goals}, from strongly correlated quantum
materials in condensed matter physics~\cite{tarruell2018quantum} to
quantum field theories in high-energy
physics~\cite{banuls2020simulating}, or modeling of complex molecules
and their dynamics in quantum
chemistry~\cite{mcardle2020quantum,arguello2019analogue}. Building an
analog quantum simulator amounts to realizing in the laboratory
synthetic, programmable quantum matter as an isolated quantum
system. Here, first of all, a specified many-body Hamiltonian $H$ must
be implemented faithfully in highly controllable quantum system with
given physical resources.  Furthermore, quantum states of matter must
be prepared on the physical quantum device corresponding to
equilibrium phases, e.g.~as ground states, or represent
non-equilibrium phenomena as in quench dynamics.

Remarkable progress has been made recently in building analog quantum
simulators to emulate quantum many-body systems. Examples are the
realization of lattice spin-models with trapped ions~\cite{Kokail2019,monroe2019programmable}, Rydberg tweezer arrays~\cite{browaeys2020many,morgado2020quantum,ebadi2020quantum},
superconducting devices~\cite{huang2020superconducting}, or Hubbard
models with ultracold bosonic or fermionic atoms in optical lattices~\cite{tarruell2018quantum,mazurenko2017cold,schafer2020tools}. While
analog quantum simulation can be viewed as special purpose quantum
computing with the rather focused task of emulating a many-body
systems via a specified $H$, the unique experimental feature is the
ability to scale to rather large particle numbers. This is in contrast
to present day quantum computers, which provide a high-fidelity
universal gate set for a small number of qubits.

Today's ability of analog quantum simulators to prepare and store on a
scalable quantum device a highly entangled many-body state, while
solving a quantum problem of physical relevance, fulfills one of the
original visions of Feynman's proposal of quantum simulation. However,
this also raises the question of verification in regimes where
comparison with classical computations with controlled error, such as
tensor network techniques, are no longer available. This includes also
higher dimensional lattice models, or with fermionic particles, and
quench dynamics.

The proper functioning of a quantum simulator can be assured by
comparing experiment vs.~theory~\cite{hauke2012can}, or predictions
from two different experimental quantum devices. This can be done on
the level of comparing expectation values of relevant observables,
e.g. on the most elementary level by comparing phase diagrams~\cite{hauke2012can}, or the increasingly complex hierarchies of
correlation functions~\cite{schweigler2017experimental}. We return to
approaches of directly comparing quantum states in Sec. III below.

\textit{Verification by Hamiltonian Learning:} Instead, we will
rephrase here verification of an analog quantum simulator as comparing
the `input' Hamiltonian, specified as the design goal for the quantum
simulator, with the actual Hamiltonian realized on the physical
device. This latter, experimental Hamiltonian can be determined via
`Hamiltonian tomography', or `Hamiltonian learning', i.e.~inferring
from measurements under certain conditions the parent Hamiltonian
underlying the experimentally prepared quantum state~\cite{bairey2019learning,qi2019determining}.

\begin{figure*}[t]
  \justifying
    \center
    \includegraphics[width=0.8\linewidth]{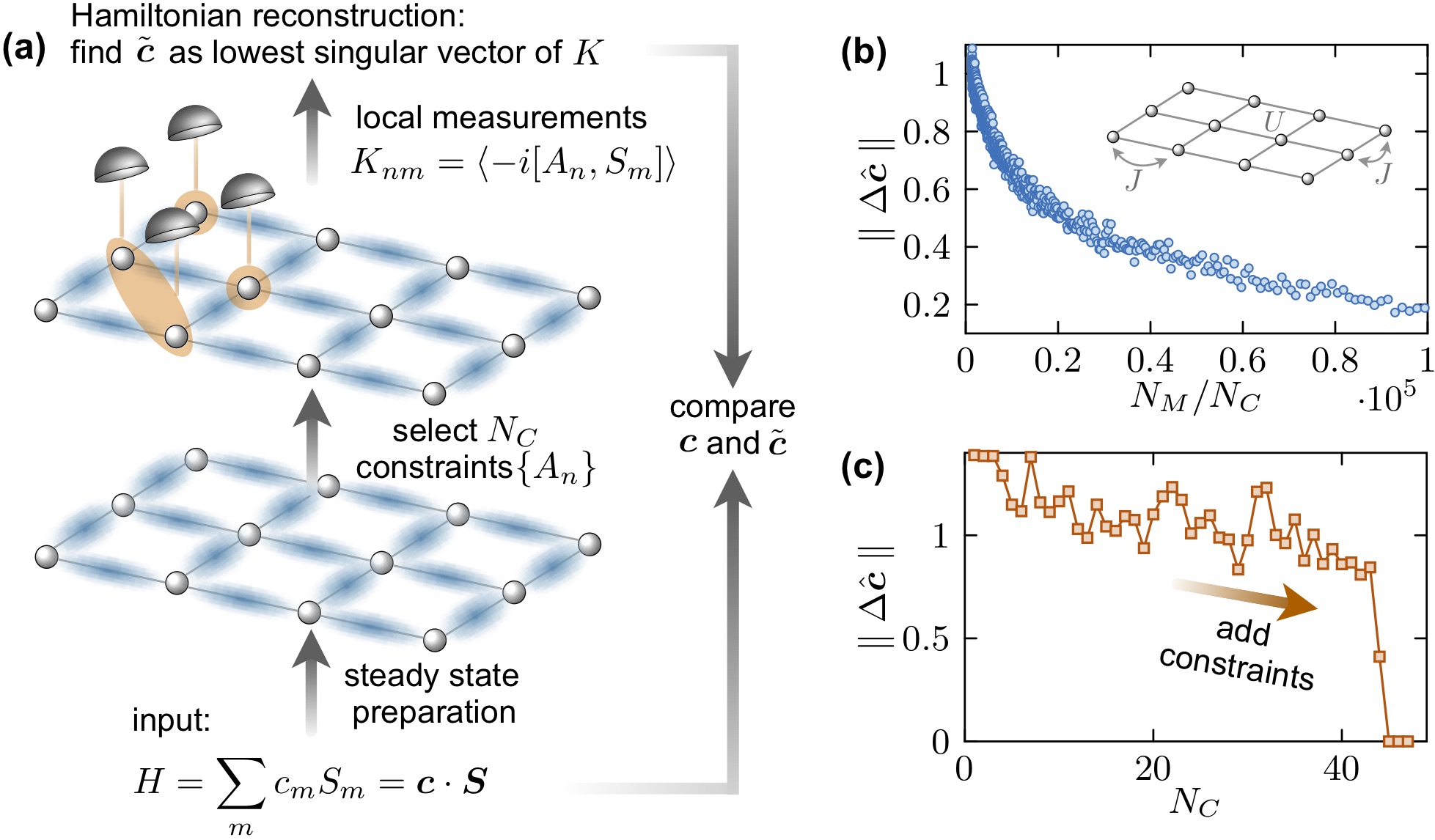}
    \hspace{0.5cm}
    \caption{(a) Schematic illustration of the protocol for verifying
      a local Hamiltonian from local measurements. (b) Number of
      required measurements (number of experimental runs) per
      constraint to achieve a certain parameter distance
      $\parallel \! \Delta \hat{\boldsymbol{c}} \! \parallel$ =
      $\parallel \! \hat{\boldsymbol{c}} -
      \tilde{\hat{\boldsymbol{c}}} \! \parallel$ for a $3\times4$
      isotropic Fermi-Hubbard model with hole-doping ($N_f=10$
      fermions). Here $\hat{\boldsymbol{c}}$ denotes the
        normalized vector
        $\hat{\boldsymbol{c}} = \boldsymbol{c}/\parallel \!
        \boldsymbol{c} \! \parallel$ with $\boldsymbol{c} = \{c_m\}$
        the expansion coefficients of the Hamiltonian $H$ in the local
        operator basis: $H = \sum_m c_m S_m$. (c) Smallest
        achievable parameter distanace
        $\parallel \! \Delta \hat{\boldsymbol{c}} \! \parallel$ as a
        function of the number of constraints $N_C$.}
    \label{fig:lindner}
    \vspace{0.5cm}
\end{figure*}

Hamiltonians of many-body physics consist of a small set of terms
which are (quasi-) \textit{local} and consist of \textit{few-body
  interactions}, i.e.~$H=\sum_i h_i$ with $h_i$ quasi-local
terms. Thus, for a given $H$, only a small set of physical parameters
determines the accessible quantum states and their entanglement
structure: for example, as ground state,
${H}\ket{\Psi_{G}}=E_{G}\ket{\Psi_{G}}$, as a finite temperature state
in the form of a Gibbs ensemble $\sim\exp{(-\beta {H})}$; or as
generator of the quench dynamics with an initial (pure) state
$\ket{\Psi_{0}}$ evolving in time as
$\ket{\Psi_{t}}=\exp{(-i{H}t)}\ket{\Psi_{0}}$.

Remarkably, as shown in recent work
\cite{bairey2019learning,qi2019determining,li2020hamiltonian}, it is
the local and few-body structure of physical Hamiltonians in operator
space which allows efficient Hamiltonian tomography via measurements
from experimentally prepared (single) quantum states on the quantum
simulator. These states include the ground state, a Gibbs state, or
states produced in quench dynamics. It is thus the restricted operator
content of Hamiltonians, which promises scalable Hamiltonian learning
with system size, i.e.~makes Hamiltonian tomography efficient.

Here we wish to outline `Hamiltonian verification' for a Fermi Hubbard
model. This can be implemented with atoms in an optical lattice, and
observed with a quantum gas microscope~\cite{tarruell2018quantum,mazurenko2017cold}. To be specific, we apply
the protocol of Ref.~\cite{bairey2019learning} for reconstruction of
the parent Hamiltonian from an experimentally prepared ground
state. Similar results apply to energy eigenstates, thermal states, or
any stationary state. We simulate experimental runs of the protocols
including the measurement budget, thus assessing accuracy and
convergence~\cite{Lorenzo2021}.

The protocol of Ref.~\cite{bairey2019learning} describes learning of
local Hamiltonians from local measurements. The starting point is the
assumption of an experimentally prepared stationary state $ \rho$, as
described above. The protocol finds the parent Hamiltonian ${H}$ from
${\rho}$ via the steady state condition $[{H}, {\rho}] = 0$. As
${\rho}$ is stationary under ${H}$, so is the expectation value of any
observable ${A}$:
$\partial_t \braket{{A}} = \braket{-i [{A}, {H}]} = 0$. The latter
equation can be used to obtain a set of linear constraints from which
${H}$ can be reconstructed. Consequently, for lattice systems the
algorithm can be summarized as follows (see also Fig.~\ref{fig:lindner}~a):
\begin{enumerate}
\item{Expand the Hamiltonian ${H}$ in a local operator basis
    ${H} = \sum_{m=1}^M c_m {S}_m$. For a $k$-local ${H}$,
    $M \sim \mathcal{O}(L^k)$ basis elements are required, with $L$
    the number of lattice site.}
\item{Select a set of $N_C > M$
    linearly independent operators (constraints)
    $\{ {A}_n \}_{n=1}^{N_C}$.}
\item{Construct a system of equations
    $\braket{-i[{A}_n, {H}]} = \sum_m c_m \braket{-i[{A}_n, {S}_m]} =
    K \boldsymbol{c} = 0$ by measuring the matrix elements
    $K_{nm} = \braket{-i[{A}_n, {S}_m]}$. }
\item{Find the lowest right singular vector $\tilde{\boldsymbol{c}}$
    of $K$, which minimizes the norm
    $\parallel \! K \boldsymbol{c} \! \parallel$.}
\end{enumerate}
As stated in Ref.~\cite{qi2019determining}, the locality of ${H}$
implies that such a Hamiltonian reconstruction will be unique.  The
reconstructed parameters $\tilde{\boldsymbol{c}}$ can be cross-checked
with respect to the parameters of an input Hamiltonian and serve as
quantifier for the verification of the quantum simulator. The required
number of experimental runs is controlled by the gap of the
correlation matrix $\mathcal{M} = K^TK$, which strongly depends on the
type and number of constraints~\cite{bairey2019learning}.  In the
limit of all possible constraints, the matrix $\mathcal{M}$ coincides
with the correlation matrix defined by Qi and Ranard~\cite{qi2019determining}. The lowest eigenvalue of this matrix
corresponds to the Hamiltonian variance measured on the input state,
which has been used previously for experimental verification of
variationally prepared many-body states~\cite{Kokail2019}.

In Fig.~\ref{fig:lindner} (b) and (c) we illustrate Hamiltonian
learning for a Fermi-Hubbard model
\begin{align}
  {H} = -J \sum_{\braket{ij}\sigma} \left( {c}_{i\sigma}^{\dagger}   {c}_{j\sigma} + \text{H.c.} \right) + U \sum_{i} n_{i \uparrow} n_{i \downarrow}
\end{align}
on a 2D square lattice~\cite{Lorenzo2021}. Here
${c}_{i \sigma}^{\dagger}$ (${c}_{i \sigma}$) denote creation
(annihilation) operators of spin-$\frac{1}{2}$ fermions at lattice
sites $i$ and ${n}_{i \sigma} =
{c}_{i\sigma}^{\dagger}{c}_{i\sigma}$. Consequently, in this
  example the local basis $\{S_m\}_{m=1}^M$ consists of hopping
  operators for all bonds $(i,j)$:
  $ ({c}_{i\sigma}^{\dagger} {c}_{j\sigma} + \text{H.c.})$ for each
  spin component $\sigma$, and of operators counting double
  occupancies on the individual sites $i$:
  $n_{i \uparrow} n_{i \downarrow}$. In case of the $3\times4$ lattice
  studied in Fig.~\ref{fig:lindner}, the operator basis therefore
  includes $M=46$ elements. As an input state for the protocol we
take the ground state in the strongly repulsive regime ($J=1$, $U=8$)
and introduce a small hole doping of $n = 0.83$. As a set of
constraints we adopt the operators
${A}_{ijk} = i ({c}_{i\sigma}^{\dagger} {c}_{j\sigma} - \text{H.c.})
n_{k \sigma'}$ in which $i$, $j$ and $k$ are nearest-neighbor
sites~\footnote{Since the ground state wave function of the Fermi
  Hubbard model as well as the Hamiltonian ${H}$ are real-valued, the
  constraints must be chosen as imaginary operators in order to obtain
  nonzero matrix elements $K_{nm} = \braket{-i[{A}_n,
    {S}_m]}$.}. The particular combinations of sites
  $\{ i,j,k \}$ is chosen in such a way that the rows of the matrix
  $K$ are linearly independent. Note that obtaining the matrix
elements $K_{nm} = \braket{-i[{A}_n, {S}_m]}$ requires the measurement
of locally resolved atomic currents
$\mathcal{J}_{i \leftrightarrow j}^{\sigma} = i
({c}_{i\sigma}^{\dagger} {c}_{j\sigma} - \text{H.c.})$, where $j$ can
be located within 2 lattice constants around $i$. In experiments with
atoms in optical lattices, these currents can be accessed by inducing
superexchange oscillations accompanied by spin resolved measurements
in a quantum gas microscope \cite{PhysRevLett.117.170405,
  PhysRevA.89.061601}.

Fig.~\ref{fig:lindner} (b) shows the relation between the distance of
the exact vs the reconstructed Hamiltonian parameters
$\parallel \! \Delta \hat{\boldsymbol{c}} \! \parallel$ and the number
of measurements per constraint on a $3\times4$ Hubbard lattice. Panel
(c) displays the improvement in quality of the Hamiltonian
reconstruction as additional constraints ${A}_{ijk}$ are added to
system of equations $K \boldsymbol{c} = 0$. As can be seen,
  the Hamiltonian can be recovered exactly as the number of
  constraints $N_C$ approaches the number of elements $M$ in the
  operator basis $\{S_m\}_{m=1}^M$. We note that the total
measurement budget can be optimized via arranging the operators
$[{A}_n, {S}_m]$ into commuting groups, such that they can be
evaluated from the same measurement outcomes \cite{Lorenzo2021}.

In the Hamiltonian learning protocol outlined above, the number of
required measurement to obtain a fixed parameter distance
$\parallel \! \Delta \hat{\boldsymbol{c}} \! \parallel$ scales
polynomially with the system size \cite{bairey2019learning}. Recent
work demonstrates that the method can be extended for recovering
Linbladians from steady states, potentially allowing an efficient
recovery of dissipative processes \cite{Bairey_2020}.  Future
investigations will have to include the relation of the type and
number of constraints to the gap of the correlation matrix which
determines the total number of required experimental runs, as well the
role of measurement errors and decoherence (see for instance
Ref.~\cite{Poggi2020}).

An entirely different verification protocol, which can also be applied
to quantum simulation, is \textit{Cross-Device Verification} described
in the following Section. There, verification is achieved by
cross-checking the results from two quantum simulators simulating the
same physics by measuring overlaps of quantum states on the level of
reduced density operators for various subsystem sizes.

\section{Cross-Device Verification of quantum computations and quantum
  simulations}
\label{sec:random_meas}

\begin{figure*}[t]
  \begin{minipage}{\linewidth}
    \center
    \includegraphics[width=0.95\linewidth]{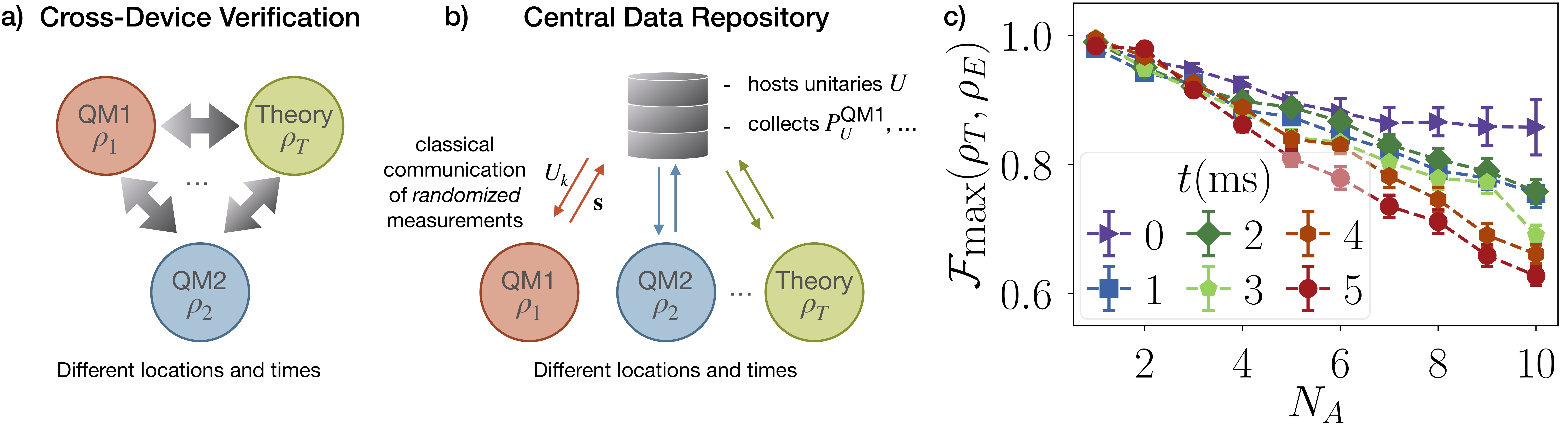} 
    \caption{(a) Cross-device verification of unknown quantum states
      $\rho_1,\rho_2,\dots$ prepared on different quantum machines
      QM1, QM2, \dots based on various platforms via (b) a central,
      standardized data-repository and classical communication of
      randomized measurements (local random unitaries and measurement
      outcomes). In regimes, where a classical simulation is possible,
      the implemented states can additionally be compared to a
      theoretical target state $\rho_T$. (c) Experiment-theory
      fidelities between a quantum states prepared in a trapped ion
      quantum simulator and its classical simulation as a function of
      the subsystem size $N_A$ (the total system consists of $10$
      qubits) for various evolution times (different colors) after a
      quantum quench in a long-range Ising model \cite{Brydges2019},
      reprinted from Ref.~\cite{Elben2020}.}
    \label{fig:xplatform}
    \vspace{0.5cm}
  \end{minipage}\\
\end{figure*}

In the previous section, we presented the verification of an analog
quantum simulator by comparing the Hamiltonian actually realized in
the device with the input or target Hamiltonian. A different approach
to verification, aiming to gain confidence into the output of a
quantum simulation or quantum computation is to run the simulation or
computation on various different quantum devices, and compare the
outcomes with each other, and - if available - with a idealized
theoretical simulation [see for an illustration
Fig.~\ref{fig:xplatform} a)]. Such cross-comparison can be implemented
on different levels of sophistication. While quantum simulations have
been compared on the level of low-order observables
\cite{hauke2012can}, for instance order parameters characterizing
phase diagrams, recent protocols aim to compare full quantum states
\cite{Flammia2011, daSilva2011,Elben2020, Huang2020}.

A convenient measure of comparison of two, possibly mixed, quantum
states $\rho_1$ and $\rho_2$ are quantum fidelities
$F(\rho_1,\rho_2)$. While various definitions of fidelities exist,
they have in common that $F(\rho_1,\rho_2)=1$ for identical states
$\rho_1=\rho_2$, and that
$F(\rho_1,\rho_2)=|\braket{\psi_1|\psi_2}|^2$ for pure states
$\rho_1=\ketbra{\psi_1}{\psi_1}$ and $\rho_2=\ketbra{\psi_2}{\psi_2}$.
A standard choice is provided by the Uhlmann fidelity
$F_U(\rho_1,\rho_2)=\trAE{\sqrt{\sqrt{\rho_1} \rho_2 \sqrt{\rho_1}}}$
\cite{Jozsa1994} which is in general difficult to access,
experimentally and also via numerical calculations
\cite{Liang2019}. Simpler alternatives are Hilbert-Schmidt fidelities
\cite{Liang2019}, such as
\begin{align}
  F_\text{max}(\rho_1,\rho_2)=\frac{\trAE{\rho_{1}\rho_{2}}}{\max\{\trAE{\rho_{1}^{2}},\trAE{\rho_{2}^{2}}\}}.
  \label{eq:Fmax}
\end{align} Here, $F_\text{max}$ defines a metric on the space of density matrices \cite{Liang2019} and coincides with the Uhlmann fidelity if one state is pure \cite{Liang2019}.

To measure quantum fidelities, various approaches exist. A pure
\emph{quantum} protocol would establish a quantum link, teleport
quantum states and compare states locally, for instance via a
SWAP-test \cite{Ekert2002,Abanin2012,Daley2012}. While such overlap
measurements have been demonstrated {locally} in seminal experiments
\cite{Islam2015,Kaufman2016,Linke2018}, a quantum link teleporting
large quantum states of many particles with high accuracy between two
quantum devices is not expected to be available in the near future.

Today, protocols relying on \emph{classical} communication between
many-body quantum devices are thus required. Here, ultimate
brute-force tests are quantum state and quantum progress tomography
which aim for a full classical reconstruction, allowing for a
  classical comparison, of quantum states or processes. Even
incorporating recent advances, such as compressed sensing for
  density matrices with low rank \cite{Gross2010}, such approach
requires however at least $3^N$ measurements to accurately determine
an arbitrary $N$ qubit states. Efficient methods, such as tensor
network \cite{Cramer2010,Lanyon2016} or neural network tomography
\cite{Torlai2018}, have been developed, rely however on a special
structure of the states of interest.

A direct approach towards fidelity estimation is provided by
randomized measurements \cite{VanEnk2012,
  Elben2018,Brydges2019,Elben2019,Elben2020, Huang2020, Ketterer2020,
  Knips_2020}. Here, a randomized measurement on a $N$--qubit quantum
state $\rho$ is performed by the application of a unitary $U$, chosen
at random from a tomographically complete set and a subsequent
measurement in the computational basis
$\left\{ \ket{\mathbf{s}} \right \}$. Statistical correlations of such
randomized measurements, performed sequentially on a single quantum
device, allow for tomographic reconstruction of the quantum state
\cite{Ohliger2013,Elben2019, Huang2020}, but also give direct access
to non-local and non-linear (polynomial) functionals of density
matrices such as R\'{e}nyi entropies \cite{VanEnk2012, Elben2018,
  Huang2020}. In particular, recent work \cite{Huang2020} combined
randomized measurements with the notion of shadow tomography \cite{Aaronson2018} which aims to predict directly expectation
  values of arbitrary observables, instead of reconstructing the full
  density matrix. Using insights from the stabilizer formalism
  \cite{Gottesman1997}, Ref.~\cite{Huang2020} devised an efficient
  implementation of shadow tomography via randomized measurements,
  which enables to estimate expectation values of arbitrary
  (multi-copy) observables with high precision and rigorous
  performance guarantees \cite{Huang2020}. This allows in particular
to estimate the fidelity between the quantum state $\rho$ and a known
theoretical target. It complements methods such as direct fidelity
estimation \cite{daSilva2011,Flammia2011} and randomized benchmarking
\cite{Emerson2005,Dankert2009,Knill2008,Emerson2007,Magesan2012,erhard2019characterizing},
which utilize the absolute knowledge of the theoretical target to be
efficient for certain target states and processes.

\textit{Cross-device verification with randomized measurements}: In a
very general setting, one faces the situation where two \emph{unknwon}
quantum states have been prepared on two separate quantum devices,
potentially at very different points in space and time
[Fig.~\ref{fig:xplatform} a)].  In Ref.~\cite{Elben2020} (see also
Ref.~\cite{Flammia2020}), it has been proposed to measure the
cross-device fidelity $F_{\text{max}}(\rho_1,\rho_2)$ of two unknown
quantum states, described by (reduced) density matrices $\rho_1$ and
$\rho_2$ and prepared on two separate devices. To this end, randomized
measurements are implemented with the \emph{same} random unitaries $U$
on both devices. Facilitating the direct experimental realization,
these unitaries $U$ can be local, $U=\bigotimes_{k=1}^{N}U_{k}$, with
$U_k$, acting on qubit $k$ and sampled from a unitary $2$-design
\cite{Gross2007,Dankert2009} defined on the local Hilbert space
$\mathbb{C}^2$. From statistical cross- and auto-correlations of the
outcome probabilities
$P_{U}^{(i)}(\mathbf{s})={\rm Tr}
[{U\,\rho_{i}\,U^{\dagger}\ket{\mathbf{s}}\bra{\mathbf{s}}}]$ of the
randomized measurements, the overlap $\trAE[]{\rho_{1}\rho_{2}}$ and
purities $\trAE[]{\rho_{1}^2}$ ($\trAE[]{\rho_{2}^2}$), and thus
$F_{\text{max}}(\rho_1,\rho_2)$, are estimated via
\begin{align}
  \trAE[]{\rho_{i}\rho_{j}}\!=\! \label{eq:ovl} 
  2^{N}\!\sum_{\mathbf{s},\mathbf{s}'}\!(-2)^{-\mathcal{D}[\mathbf{s},\mathbf{s}']}\;\overline{P_{U}^{(i)}(\mathbf{s})P_{U}^{(j)}(\mathbf{s}')},
\end{align}
for $i,j=1,2$.  Here, $\overline{\vphantom{a}\dots}$ denotes the
ensemble average over local random unitaries and the Hamming distance
$\mathcal{D}[\mathbf{s},\mathbf{s}']$ between two strings $\mathbf{s}$
and $\mathbf{s}'$ is defined
as 
$\mathcal{D}[\mathbf{s},\mathbf{s}']\equiv\left| \left\{ k\in
    \{1,\dots, N\}\,|\,s_{k}\neq{s}'_{k}\right\}\right|$.
		
In the regime where a classical simulation of the output is possible,
this protocol can also be used for a experiment-theory comparison
(c.f.\ direct fidelity estimation \cite{Flammia2011, daSilva2011} and
classical shadow tomography \cite{Huang2020}). In
Fig.~\ref{fig:xplatform}(c), experiment-theory fidelities between
highly entangled quantum states prepared via quench dynamics in a
trapped ion quantum simulator \cite{Brydges2019} and its theoretical
simulation are shown \cite{Elben2020}.  We note that such
experiment-theory comparisons to simple (product) states can also be
used to identify and mitigate errors resulting from imperfect
measurements \cite{Elben2020,Chen2020,Koh2020}.

Based on numerical simulations, it was found in Ref.~\cite{Elben2020}
that the number of necessary experimental runs to estimate the
fidelity $F_\text{max}$ up to a fixed statistical error scales
exponentially with subsystem size, $\sim 2^{bN}$. The exponents
$b\lesssim 1$ are however favorable compared to quantum state
tomography, enabling fidelity estimation for (sub-) systems consisting
of a few tens of qubits with state of the art quantum devices.  For
two very large quantum devices, consisting of several tens to a few
hundreds of qubits, the present protocol allows thus only to estimate
fidelities of possibly disconnected subsystems up to a given size,
determined by the available measurement budget.  This data represents
very fine-grained local information on fidelities of subsystems.  It
remains an open question whether this information can be combined with
additional knowledge of a few global properties to obtain (at least
bounds on) the total system fidelity.

While we have outlined above protocols to cross-check two individual
devices, we envision a community effort where specific quantum
problems, either as quantum circuits and algorithms, or for quantum
simulation are defined, and data from theoretical simulations as well
as measurement data from quantum devices are uploaded to a central
data repository [see for an illustration Fig.~\ref{fig:xplatform}
b)]. In regimes, where a classical simulation is possible, an ultimate
reference could here be represented by a theory target state. For
larger quantum devices, reference operations and circuits could be
executed, and density matrices of (sub-)systems could be compared with
each other. This would allow for a standardized, pairwise cross-check
of multiple quantum devices representing various platforms.

The outlined protocols rely on classical communication of randomized
measurement results, and are restricted, due to an exponential scaling
of the number of required experimental runs, to (sub)-systems of a few
tens of qubits. To overcome this challenge, we expect that in the
future quantum state transfer protocols become available to develop
efficient fully quantum in addition to hybrid quantum-classical
protocols for cross-checking quantum devices.

\section{Verification of the output of an untrusted quantum device}

In the validation procedures considered above, the person testing the
quantum processor (the user) has either direct access to the device or
trusts the person operating it. Computer scientists are often
concerned about a very different notion of verification: \textit{the
  verification of the output of a computation performed by an
  untrusted device}. Such a verifiability demand will become
particularly relevant once quantum devices, that reliably process
hundreds of qubits, become usable as cloud computers.

To demonstrate the need of these verification protocols, let us
consider the various kinds of problems such cloud computers could be
utilized for. If the user employs a quantum computer to solve a
problem within NP, such as factoring a large number into its prime
factors, the solution to the problem is simple: knowing the factors,
the output can be efficiently verified with a classical
computer. However, it is believed that quantum computers are capable
of efficiently solving problems that can no longer be efficiently
verified classically, such as simulating quantum many-body systems.
How can one then rely on the output, given that the quantum computer
(or the person operating it) might be malicious and wants to convince
the user that the answer to e.g. a decision problem is "yes" when it
is actually "no"?  Hence, harnessing the full power of a quantum
device, which is not directly accessible to the user, brings with it
the necessity to derive protocols for verifying its output. The aim
here is to derive quantum verification protocols that allow a
computationally limited (e.g. a classical) user to verify the output
of a (powerful) quantum computer.  Complicating matters, is the need
to ensure that a honest prover (the quantum computer) can convince the
user of the correct outcome efficiently~\cite{AV12}. To express things
more simply, we will refer now to the user (called verifier) as Alice
($\sf A$) and to the prover as Bob ($\sf B$).

Verification protocols~\footnote{For an excellent, recent review of
  these protocols see~\cite{GhKa19}.} where $\sf A$ has access to
limited quantum
resources~\cite{MF13,Mo14,HM15,GKP15,HPF15,FK17,ABEM17,TMH19} or is
able to interact with two non--communicating provers have been
derived~\cite{RUV12}. In a recent breakthrough Mahadev~\cite{Ma18}
showed that even a purely classical user can verify the output of a
quantum processor. In contrast to the verification protocols mentioned
before, this protocol relies on a computational assumption: the
existence of trapdoor functions which are post-quantum
secure~\footnote{These functions are constructed
  in~\cite{Ma18,Ma18IEEE} based on the hardness of the Learning With
  Errors (LWE) problem~\cite{Re05}}. These functions are hard to
invert even for a quantum computer. However the possession of
additional information (trapdoor) enables one to compute the preimages
of the function efficiently. Using the notion of post-quantum secure
trapdoor functions in combination with powerful previously derived
findings, led to the surprising result that a classical user can
indeed verify the output of a quantum computer, as we will briefly
explain below. The notion and techniques developed
in~\cite{Ma18,Ma18IEEE} have recently been utilized to put forward
protocols with e.g.  zero-knowledge polynomial-time
verifiers~\cite{VZ20} and non-interactive classical
verification~\cite{ACGH20}.

At a first glance it simply seems impossible to efficiently verify the
output of a much more powerful device (even if it was classical) if
one is just given that output and is prevented from testing the
device. The key idea here is to use interactive proofs. The exchange
of messages allows $\sf A$ to test $\sf B$ and to eventually get
convinced that the $\sf B$'s claim is indeed correct or to mistrust
him and reject the answer. The Graph Non--isomorphism problem is a
simple example of a task where the output (of a powerful classical
device) can be verified with an interactive proof~\cite{Vi19}.

\begin{figure*}[t]
  \begin{minipage}{\linewidth}
    \centering \includegraphics[width=\linewidth]{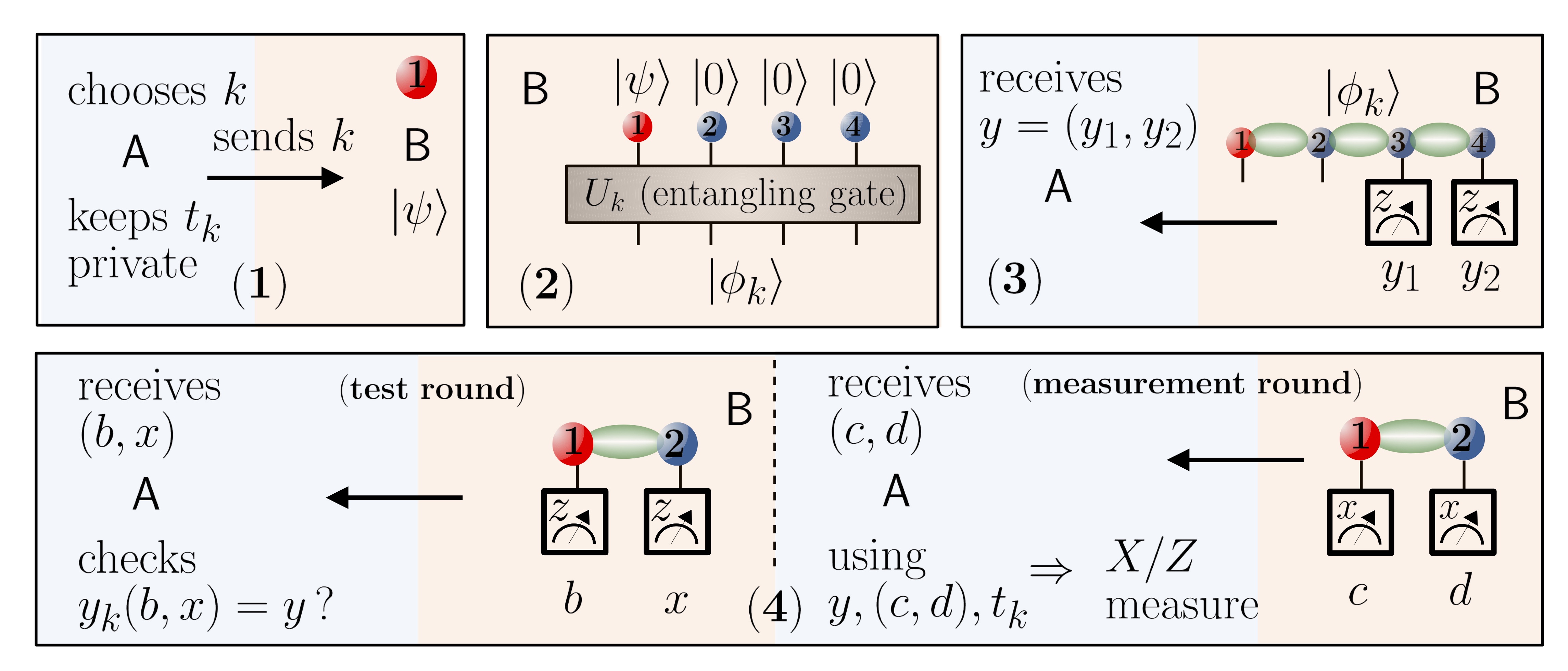}
    \caption{Sketch of steps 1--4 in the measurement protocol
      \cite{Ma18} (see main text). Alice ($\sf A$) represents a
      classical verifier, Bob ($\sf B$) an efficient quantum prover,
      and we have used arrows to emphasise the flow of classical
      information. $\sf A$ performs a delegated measurement on a
      single-qubit state $\ket\psi=\alpha_0\ket0+\alpha_1\ket1$ in
      either the $Z$- or $X$-basis without revealing the measurement
      basis to $\sf B$. An appropriate combination of the statistics
      of the results of such measurements allows $\sf A$ to calculate
      the energy of the state prepared by $\sf B$ and thereby verify
      the outcome of the original decision problem.}
    \label{fig:mp}
  \end{minipage}
\end{figure*}

To explain the general idea of how to verify the output of a quantum
device, we assume that $\sf B$ possesses a quantum computer, whereas
$\sf A$ only has classical computational power. $\sf A$ asks $\sf B$
to solve a decision problem (within BQP, i.e. a problem which can be
solved efficiently by a quantum computer) and wants to verify the
answer. Of particular importance here is that one can show that the
outcome of such a decision problem can be encoded in the ground state
energy of a suitable, efficiently computable, local Hamiltonian
$H$~\cite{KKR05}. This implies that in case $\sf B$ claims that the
answer to the decision problem is "yes"~\footnote{{The case in which
    $\sf B$ claims that the answer to the decision problem is "no" can
    be treated similarly.}}, he can convince $\sf A$ of this fact by
preparing a state with energy (w.r.t. $H$) below a certain value,
which would be impossible in case the correct answer was "no". An
instance of such a state is the so--called clock--state,
$\ket{\eta}$~\cite{Fe86,KSV02}, which can be prepared efficiently by a
quantum computer. Hence, the output of the quantum computer can be
verified by determining the energy of the state prepared by $\sf
B$. This can be achieved by performing only measurements in the $X$--
as well as the $Z$--basis~\cite{BL08,MNS16,FHM18}. It remains to
ensure that $\sf A$ can delegate these measurements to $\sf B$ without
revealing the measurement basis. The important contribution of
Mahadev~\cite{Ma18} is the derivation of such a measurement protocol
(see Fig.~\ref{fig:mp}). The properties of post--quantum secure
trapdoor functions are exploited precisely at this point to ensure
that $\sf B$ can not learn whether a qubit is measured in the $Z$- or
$X$-basis, which prevents him from cheating.

For reasonable choices of the security parameters, the realization of
such a verification protocol is, even without considering faulty
devices, not feasible with current technology (on $\sf B$'s
side). Already the number of auxiliary qubits required in the
measurement protocol would be too demanding (\cite{Li11} and~\cite{BrCh}). Nevertheless, due to the rapid technological
development and the accompanying need for these kind of verification
protocols, we present here a proposal for a proof--of--principle
experiment. The minimal example explained here can already be carried
out with a total of seven qubits.

First, the original decision problem is mapped to a local Hamiltonian
problem. $\sf B$ prepares the corresponding state, $\ket{\eta}$
(consisting of 4 qubits in this example), whose energy needs to be
determined. Due to the linearity of the measurement protocol, it is
sufficient to demonstrate how $\sf A$ can delegate the measurement in
the $X$- or $Z$-basis on a single-qubit state
$\ket\psi=\alpha_0\ket0+\alpha_1\ket1$ (belonging to $\ket{\eta}$)
without revealing the measurement basis to $\sf B$.  The measurement
protocol has the following high-level structure (see
Fig.~\ref{fig:mp}):

\begin{enumerate}
\item $\sf A$ computes a family of post--quantum secure trapdoor
  functions $\{y_k\}$, labeled by an index $k$, together with the
  associated information $t_k$ (trapdoor). The functions $y_k$ are of
  one of the two types, either one-to-one or
  two-to-one~\footnote{Note that the two-to-one functions
      are such that $y_k(0,\cdot)$ and $y_k(1,\cdot)$ are
      injective}. If $\sf A$ wants to measure $\ket\psi$ in the
  $Z$-basis ($X$-basis), she chooses a label $k$, such that $y_k$ is
  one-to-one (two-to-one), respectively. $\sf A$ keeps $t_k$ private
  (this is precisely the leverage $\sf A$ has over $\sf B$) and sends
  $k$ to $\sf B$.  Knowing $k$, $\sf B$ can efficiently evaluate the
  function $y_k$ on any input. However, it is computationally hard for
  him to determine which type $y_k$ is. Furthermore, $\sf A$ can
  compute the preimages of $y_k$ efficiently using $t_k$ while $\sf B$
  cannot.

\item $\sf B$ is asked to prepare the state
  $\ket{\phi_k}\propto\sum_{b,x}\alpha_b\ket b\ket
  x\ket{y_k(b,x)}$. This can be done efficiently by a quantum
  computer.

\item $\sf B$ is asked to measure the last register (qubits 3 and 4 in
  our example) of $\ket{\phi_k}$ in the $Z$-basis and to send the
  measurement outcome $y$ to $\sf A$. The state of the first and
  second register (qubits 1 and 2 in our example) is then, depending
  on the type of $y_k$ either: (i) The product state $\ket b\ket x$
  (with probability $|\alpha_b|^2$) where $y_k(b,x)=y$; or (ii) The
  entangled state $\alpha_0\ket0\ket{x_0}+\alpha_1\ket1\ket{x_1}$
  where $y_k(0,x_0)=y_k(1,x_1)=y$.

\item $\sf A$ randomly chooses to run either a "test" or a
  "measurement" round. In a "test" ("measurement") round, $\sf B$ is
  asked to measure the qubits in the first and second register in the
  $Z$-basis ($X$-basis) respectively, and to send the outcome to
  $\sf A$. The "test" rounds allow $\sf A$ to gain confidence that
  $\sf B$ has indeed prepared $\ket{\phi_k}$ by checking that
  $y_k(b,x)=y$. In a "measurement" round the first qubit has
  effectively been measured in either the $Z$ or $X$ basis, depending
  on the type of $y_k$. Using the trapdoor information $t_k$, $\sf A$
  can classically post-process the outputs to obtain the corresponding
  measurement outcome.
\end{enumerate}

As mentioned above, a minimal, non-trivial example, which can be
realized with an ion-trap quantum computer~\cite{Cnp} requires only
$7$ qubits in total and some tens of single and two--qubit
gates~\footnote{This would correspond to a decision problem encoded in
  the output of a quantum circuit composed of three single- and
  two-qubit gates acting on two qubits initialized in the state
  $\ket0^{\otimes2}$.}. In this case the clock-state $\ket\eta$ is a
$4$-qubits state and, for this minimal example, one can choose the
second and third register to have $1$ and $2$ qubits, respectively (as
displayed in Fig.~\ref{fig:mp}). Here, $y_k:\{0,1\}^2\to\{0,1\}^2$ and
$k$ labels either one of the $24$ one-to-one functions or
  one of the $24$ two-to-one functions.

Let us finally mention that protocols which allow the verification of
the output of imperfect quantum computers have been recently put
forward in case the verifier has limited access to quantum
resources~\cite{GhHo19}. Similar ideas can also be utilized in the
purely classical verification protocol~\cite{Ma18}, ensuring that the
measurements can still be performed without jeopardizing its
security~\cite{Cnp}.

\section{Conclusion and Outlook }

In an era, where we build noisy intermediate scale quantum devices,
with the effort to scale them towards larger system sizes and optimize
their performance, verification of quantum devices becomes a main
motif in theoretical and experimental quantum information science. In
this perspective on theoretical and experimental aspects of quantum
verification we have taken an approach of discussing three examples,
formulated as proposed experiments. The three examples are
verification of quantum simulation via Hamiltonian learning (Sec.~II),
cross-checking of quantum states prepared on different quantum devices
(Sec.~III), and addressing the question of how a user of a quantum
processor can be certain about the correctness of its output
(Sec.~IV).  While our choice of examples highlighting quantum
verification is subjective and guided by personal interests, the
common theme is that these `proposed experiments' can be performed
with the quantum devices existing in today's quantum laboratories or
near-future devices. In addition, our examples illustrate the
diversity of questions in quantum verification and tools and
techniques to address them, with emphasis on what we identify as
problems of high relevance.

Of course, by the very nature of a perspective as forward looking, we
identify interesting topics and outline possible avenues, while
putting the finger on open issues for future theoretical and
experimental work. These open problems range from technical to
conceptual issues, and we summarize some of these questions within the
various sections. The overarching challenge is, of course, to develop
efficient and quantitative verification protocols and techniques,
which eventually scale to large system sizes we envision as useful
quantum devices. In Sec.~II on verification of analog quantum
simulation via Hamiltonian learning, the local Hamiltonian ansatz
scales, by construction, with the system size, and leads, in
principle, to a quantified error assessment. While one may raise
issues of imperfect state preparation and measurement errors in
experiments, and the measurement budget available in a specific
experiment, we emphasize that these protocols also involve heavy
classical post-processing of data, which may provide limits from a
practical and conceptual perspective. While this might not pose
serious limitations for near-term devices, we might ask here, but also
in a broader context, if some of this post-processing can be replaced
by more efficient quantum post-processing on the device. The
cross-device check of quantum states in Sec.~III provides another
example of this type. There, the protocol underlying the comparison of
quantum states with a central data repository involves classical
communication. The protocol described is much more efficient than
tomography, and scales with a `friendly exponential' in system size,
allowing today experimental implementation for tens of qubits. A
future development of quantum state transfer as quantum communication
between the devices promises to overcome these limitations. Finally,
our discussion in Sec.~IV on verification of the output of an
untrusted quantum device presents an absolute minimal example which
can be run on present quantum computers, leaving as challenges the
verification of outputs of imperfect quantum devices and more advanced
experimental demonstrations.

Verification of quantum processors become particularly challenging and
relevant in the regimes of quantum advantage, where quantum devices
outperform their classical counterparts~\cite{Go19,Zh20}. As solving a
"useful" computational task (such as factoring a large number) would
neither be feasible with noisy intermediate--scale quantum computers
nor necessary to demonstrate quantum superiority, one focuses on
sampling problems~\cite{AA10,Go18,BFNV19}, in this context. However,
these approaches entail difficulties in demonstrating quantum
superiority. On the one hand, the fact that the sampling was performed
faithfully needs to be verified. On the other hand, one needs to show
that the task is computationally hard for any classical device (taking
into account that the quantum computer is imperfect). In this context,
both, strong complexity--theoretical evidence of classical
intractability as well as new proposals for experimental realizations
for various setups are desirable.

\textit{Acknowledgment -- } Work at Innsbruck is supported by the
European Union program Horizon 2020 under Grants Agreement No.~817482
(PASQuanS) and No.~731473 (QuantERA via QTFLAG), the US Air Force
Office of Scientific Research (AFOSR) via IOE Grant
No.~FA9550-19-1-7044 LASCEM, by the Simons Collaboration on
Ultra-Quantum Matter, which is a grant from the Simons Foundation
(651440, PZ), by the Austrian Science Fund (FWF): stand alone project
P32273-N27, by the SFB BeyondC, and by the Institut f\"ur
Quanteninformation.

\bibliography{Bib_Verification}

\end{document}